\documentclass[aps,prc,twocolumn,reprint,floatfix,superscriptaddress,showpacs]{revtex4-1}
\usepackage{graphicx}
\usepackage[pdftex,colorlinks,citecolor=blue,bookmarks]{hyperref}
\usepackage{bm}
\usepackage{longtable}
\begin{document}
\title{Octupole correlations in the nucleus $^{144}$Ba described with symmetry conserving configuration mixing calculations}
\author{R\'emi N. Bernard} 
\author{Luis M. Robledo} 
\author{Tom\'as R. Rodr\'iguez} 
\affiliation{Departamento de F\'isica Te\'orica, Universidad Aut\'onoma de Madrid, E-28049 Madrid, Spain}
\begin{abstract}
We study the interplay of quadrupole and octupole degrees of freedom in the structure of the isotope $^{144}$Ba. A symmetry conserving configuration mixing method (SCCM) based on a Gogny energy density functional (EDF) has been used. The method includes particle number, parity and angular momentum restoration as well as axial quadrupole and octupole shape mixing within the generator coordinate method. Predictions both for excitation energies and electromagnetic transition probabilities are in good agreement with the most recent experimental data.   
\end{abstract}
\maketitle
Permanent octupole deformation is a rare phenomenon in atomic nuclei produced by the octupole interaction between two opposite parity single particle levels with $\Delta l=\Delta j=3$ near the Fermi surface~\cite{RMP_68_349_1996}. The distribution of single particle levels for a certain number of particles such as 56, 88, 134, etc. favor such an interaction. As a consequence, when both protons and neutron numbers are close to these values, strong octupole correlations are expected. This happens, for instance, in heavy nuclei around $^{220}$Rn and $^{224}$Ra~\cite{Nature_497_199_2013} or in medium-mass nuclei around $^{144}$Ba where low-lying negative parity states have been found experimentally as an indication of this kind of correlations. In fact, recent multi-step Coulomb excitation experiments performed at the ATLAS-CARIBU facility with state-of-the-art $\gamma$-ray (GRETINA) and charged-particle (CHICO2) detectors have shown large $E3$ transitions that evidence permanent octupole deformation in the isotope $^{144}$Ba~\cite{PRL_116_112503}.

The study of octupole correlations and the associated breaking of the reflection symmetry is still a challenge for nuclear theory. Microscopic self-consistent mean field (MF) methods~\cite{RMP_75_121_2003} based on nuclear energy density functionals (EDF) such as Skyrme, Gogny and/or Relativistic Mean Field (RMF) are always the starting point as they have been largely improved in the last fifteen years by including beyond-mean-field (BMF) correlations. In particular, symmetry restorations and mixing of different mean-field many-body states have been implemented within the general framework provided by the generator coordinate method (GCM)~\cite{RingSchuck}. These developments have allowed to study the impact of octupole correlations  not only in ground state properties, such as binding energies and radii, but also in nuclear spectra, electromagnetic transitions and decays all over the periodic table with Skyrme or Gogny functionals \cite{PRL_66_876_1991,NPA_524_65_1991,NPA_551_109_1993,NPA_559_221_1993,NPA_574_185_1994,PRC_50_802_1994,NPA_588_597_1995,PRL_80_4398_1998,PRC_84_054302_2011,PRC_88_051302R_2013,JPG_42_055109_2015}. An alternative to MF methods is the extension of the IBM to include negative parity bosons \cite{engel87} to handle negative parity states. A nice reproduction of experimental data is obtained, see \cite{kuszenov88,nomura15} as examples, but at the cost of introducing several adjustable parameters.

On the other hand, the quadrupole degree of freedom together with pairing play a dominant role in describing low energy nuclear correlations. Hence, the restoration of the associated broken symmetries, i.e., particle number and angular momentum, has been implemented with different levels of complexity including  axial~\cite{NPA_671_145_2000,PRL_99_062501_2007,PRC_74_064309_2006} and non-axial~\cite{PRC_78_024309_2008,PRC_81_064323_2010,PRC_81_044311_2010} quadrupole deformed intrinsic states. Additionally, other degrees of freedom such as pairing fluctuations~\cite{PLB_704_520_2011,PRC_88_064311_2013} and/or intrinsic (cranking) rotational frequencies~\cite{PLB_746_341_2015,PRL_116_052502_2016} have been studied in combination with the quadrupole deformation. These symmetry conserving configuration mixing methods (SCCM) show a nice performance in describing qualitatively nuclear structure phenomena like appearance/degradation of shell closures, shape coexistence, high spin physics, etc. 

Octupole shapes have been scarcely explored within the angular momentum and particle number projected SCCM framework despite the very likely coupling to quadrupole and pairing degrees of freedom. Only very recently, GCM calculations with simultaneous particle number, parity and angular momentum projected states have been reported within the RMF framework~\cite{PRC_92_041304_2015,PLB_753_227_2016} to study octupole excitations in $^{224}$Ra and clustering of $^{20}$Ne. In this work, we report on the first implementation of this scheme (particle number, parity and angular momentum projection plus GCM) with the Gogny EDF and its application to the study of the lowest positive and negative parity states of the nucleus $^{144}$Ba.


Nuclear states with angular momentum and parity quantum numbers $J^{\pi}$ are obtained within the present SCCM method through the GCM ansatz~\cite{RingSchuck}
\begin{equation}
|\Psi^{J\pi}_{\sigma}\rangle=\sum_{\mathbf{q}}f^{J\pi}_{\sigma} (\mathbf{q})|\Phi^{J\pi} (\mathbf{q})\rangle
\label{GCM_ansatz} 
\end{equation}
where $\sigma=1,2,...$ labels the different quantum states for a given $J^{\pi}$ and $|\Phi^{J\pi} (\mathbf{q})\rangle$ are the projected intrinsic states
\begin{equation}
|\Phi^{J\pi} (\mathbf{q})\rangle=P^{J}P^{\pi}P^{N}P^{Z}|\mathbf{q}\rangle
\end{equation}
with $P^{J}$, $P^{\pi}$, $P^{N}$ and $P^{Z}$ being the projectors onto good angular momentum, parity, neutron number and proton number respectively~\cite{RingSchuck}. Furthermore, the intrinsic states, $|\mathbf{q}\rangle$, are obtained by solving Hartree-Fock-Bogoliubov (HFB) equations, imposing the constraints on the corresponding collective coordinates $\mathbf{q}=\{q_i,i=1,\ldots,N_c\}$ \cite{RingSchuck}.


In the present work, axial quadrupole and octupole degrees of freedom ($\mathbf{q}=(q_{20},q_{30})$), or equivalently, $(\beta_{2},\beta_{3})$, are explored explicitly. The dimensionless $\beta_\lambda$ parameters are defined as
$
\beta_{\lambda}\equiv{4\pi}\langle\mathbf{q}\,|r^{\lambda}Y_{\lambda0}|\mathbf{q}\rangle / ({3r^{\lambda}_{0}A^{\lambda/3+1}})
$
with $r_{0}=1.2$ fm and $A$ being the mass number.

We impose axial, time-reversal and simplex symmetries in the HFB wave functions because these conditions significantly reduce  the computational cost of the calculations. In particular, two of the three integrals in the Euler angles required to perform the angular momentum projection are trivial in this case. However, this choice presents some restrictions. On the one hand, triaxial nuclear states are out of the present description. Moreover, the accessible states are limited to those satisfying the $(-1)^J=\pi$ rule (even $J$ with positive parity and odd $J$ with negative parity). Finally, a systematic stretching of the spectrum is expected because this kind of variational method favors the ground state energy over the energies of the excited states~\cite{Rod.00,PLB_746_341_2015}. 

The coefficients of the linear combination given in Eq.~\ref{GCM_ansatz} are found by solving the so-called Hill-Wheeler-Griffin (HWG) equations, one for each value of the angular momentum and parity~\cite{RingSchuck}
\begin{equation}
\sum_{\mathbf{q}'} \left(\mathcal{H}^{J\pi} (\mathbf{q},\mathbf{q} ')-E^{J\pi}_{\sigma}\mathcal{N}^{J\pi} (\mathbf{q}, \mathbf{q}')\right) f^{J\pi}_{\sigma} (\mathbf{q}')=0\label{HWG_1}
\end{equation}
with the norm $\mathcal{N}^{J\pi} (\mathbf{q}, \mathbf{q}')=\langle\Phi^{J\pi} (\mathbf{q})|\Phi^{J\pi} (\mathbf{q}')\rangle$ and Hamiltonian $\mathcal{H}^{J\pi} (\mathbf{q},\mathbf{q} ')=\langle\Phi^{J\pi} (\mathbf{q})|\hat{H}|\Phi^{J\pi} (\mathbf{q} ')\rangle$ overlaps. Given the nature of the density-dependent term of the Gogny EDF a prescription is required for the evaluation of Hamiltonian overlaps. We use the particle number projected spatial density combined with the mixed prescription for the parity and angular momentum projection and GCM parts.  It avoids the catastrophic behavior of the energy characteristic of prescriptions based on densities preserving spatial symmetries \cite{Rob.10}. The impact of the use of the particle number projected density has still to be elucidated.
\begin{figure}[t]
\begin{center}
\includegraphics[width=\columnwidth]{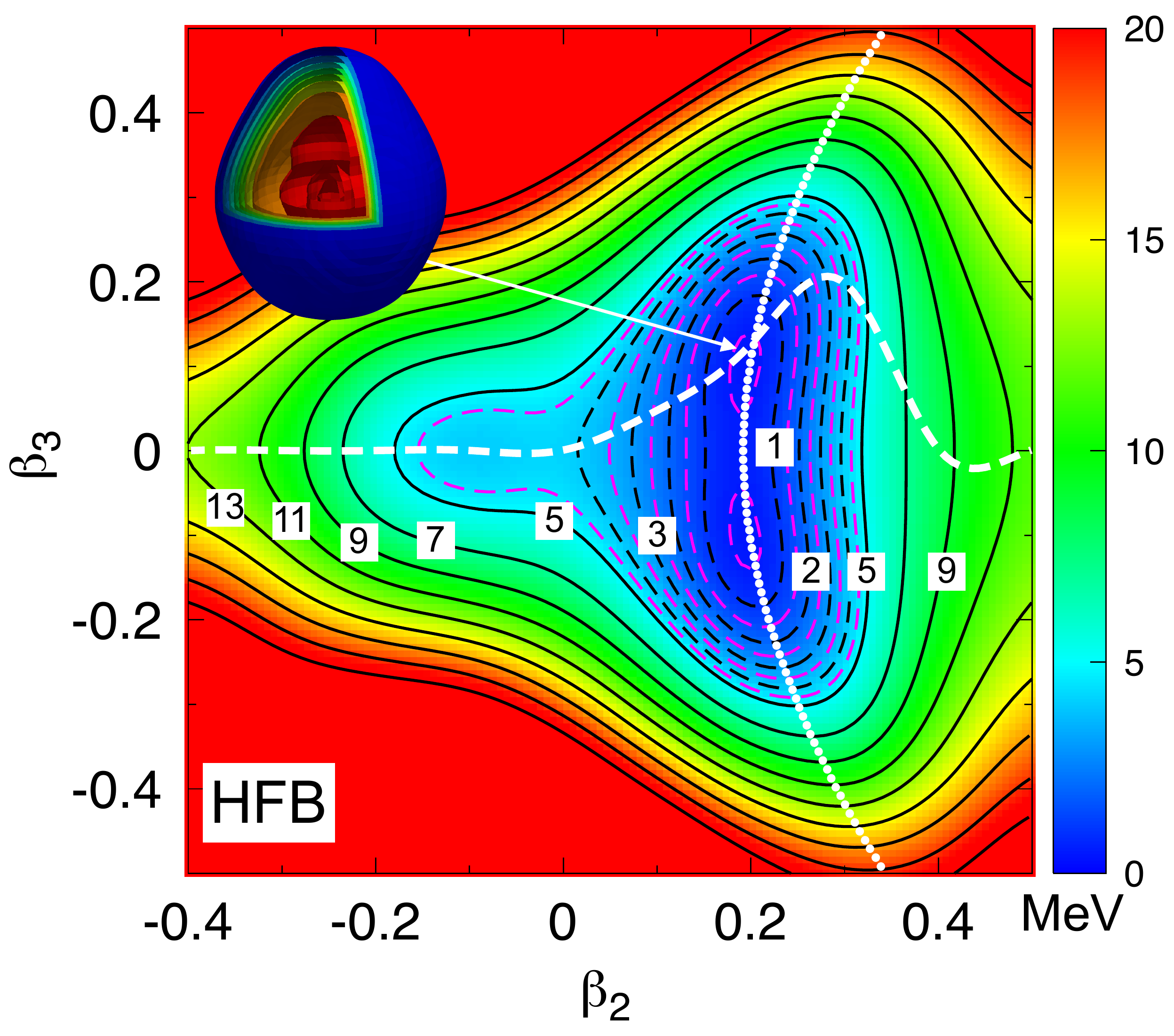}
\end{center}
\caption{(color online) HFB potential energy surface in the $(\beta_{2},\beta_{3})$ plane normalized to the energy of the minimum ($-1180.772$ MeV with eleven HO shells) computed for $^{144}$Ba with the Gogny D1S parametrization. Contour lines are separated by 0.5 MeV (dashed lines) and 2.0 MeV (full lines) respectively. White dashed (dotted) lines represent the paths for one-dimensional $\beta_{2}$ ($\beta_{3}$) constrained  HFB calculations. The spatial density of the HFB ground state is also plotted.}
\label{Fig1}
\end{figure}
\begin{figure}[b]
\begin{center}
\includegraphics[width=\columnwidth]{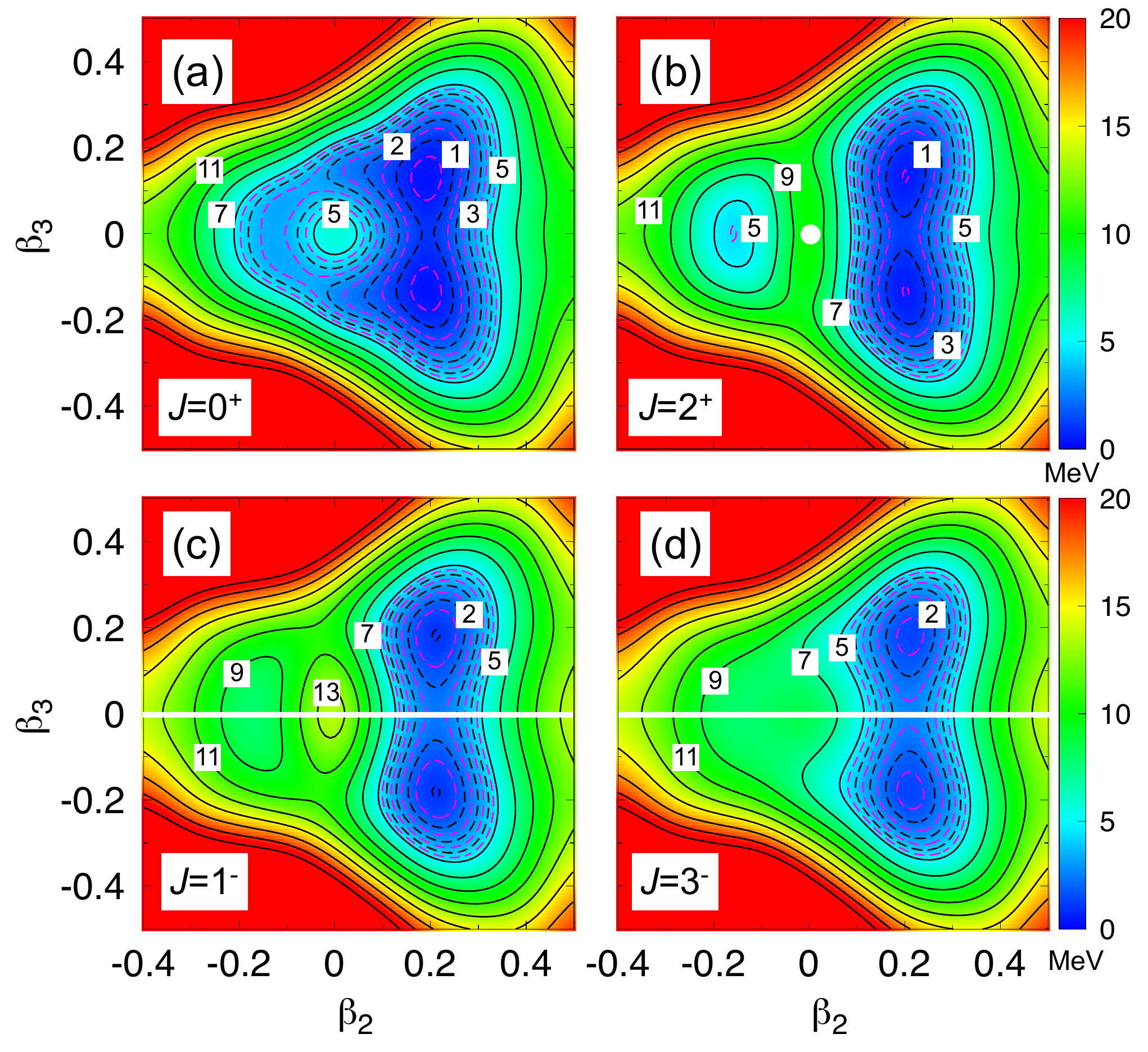}
\end{center}
\caption{(color online) Particle number, parity and angular momentum PES in the $(\beta_{2},\beta_{3})$ plane normalized to the energy of the minimum of the $0^{+}$-PES ($-1185.600$ MeV with eleven HO shells). Contour lines are separated by 0.5 MeV (dashed lines) and 2.0 MeV (full lines) respectively.}. 
\label{Fig2}
\end{figure}

The HWG generalized eigenvalue problem is routinely solved by transforming it into a regular one by introducing a set of orthonormal states -the natural basis- defined as linear combinations of the non-orthonormal states, $\lbrace|\phi^{J\pi}(\mathbf{q})\rangle\rbrace$. Once the equations are solved,
the spectrum is directly given by $E^{J\pi}_{\sigma}$. Furthermore, expectation values and transition probabilities are computed from the coefficients $f^{J\pi}_{\sigma} (\mathbf{q})$ and the projected matrix elements of the corresponding operators, $\hat{O}$:
\begin{eqnarray}
\langle\Psi^{J_{1}\pi_{1}}_{\sigma_{1}}|\hat{O}|\Psi^{J_{2}\pi_{2}}_{\sigma_{2}}\rangle=&\nonumber\\
\sum_{\mathbf{q}_{1},\mathbf{q}_{2}}\left(f^{J_{1}\pi_{1}}_{\sigma_{1}} (\mathbf{q}_{1})\right)^{*}
\mathcal{O}^{J_{1}\pi_{1}, J_{2}\pi_{2}} (\mathbf{q}_{1}, \mathbf{q}_{2})\left(f^{J_{2}\pi_{2}}_{\sigma_{2}} (\mathbf{q}_{2})\right)&
\end{eqnarray}
with $\mathcal{O}^{J_{1}\pi_{1}, J_{2}\pi_{2}} (\mathbf{q}_{1}, \mathbf{q}_{2})=\langle\Phi^{J_{1}\pi_{1}} (\mathbf{q}_{1})|\hat{O}|\Phi^{J_{2}\pi_{2}} (\mathbf{q}_{2})\rangle$ the overlap of the operator $\hat{O}$, which is not necessarily a scalar operator. Finally,  the weights of the different natural basis states in a given GCM wave function~\cite{RingSchuck}:
\begin{equation}
F^{J\pi}_{\sigma} (\mathbf{q})\equiv \sum_{\mathbf{q}'} \langle\Phi^{J\pi} (\mathbf{q})|\Phi^{J\pi} (\mathbf{q}\,')\rangle^{1/2}f^{J\pi}_{\sigma} (\mathbf{q}\,')\label{coll_wf}
\end{equation}
are very useful quantities to analyze the character of the GCM states.

The method described above is now used to compute the low-lying energy spectrum and electromagnetic transition probabilities of the nucleus $^{144}$Ba. A set of 130 intrinsic HFB states in the ranges $\beta_{2}\in\left[-0.4,0.5\right]$ and $\beta_{3}\in\left[-0.5,+0.5\right]$ has been chosen. Each of the HFB wave functions are expanded in eleven spherical harmonic oscillator (HO) shells. The number of integration points in the $\beta$-Euler angle and $\varphi$-gauge angle (particle number projection) are 32 and 9 respectively. These values ensure a proper convergence of the expectation values of the particle number and total angular momentum operators between GCM states, as well as the SCCM spectrum and collective wave functions (see, e.g., Refs.~\cite{RMP_75_121_2003,NPA_709_201_2002,PRC_81_064323_2010} for more details). Finally, the Gogny D1S interaction has been  consistently used in this work both at the mean field level and beyond.

The role of the different collective degrees of freedom can be guessed by looking at the mean-field energy landscape as a function of such coordinates. In Fig.~\ref{Fig1} the potential energy surface (PES) of the nucleus $^{144}$Ba in the $\beta_{2}-\beta_{3}$ plane is shown. Notice that the energy is symmetric under a change in the sign of $\beta_{3}$ due to the parity symmetry of the nuclear interaction. Hence, two symmetric absolute minima are obtained at $(\beta_{2},\beta_{3})=(0.2, \pm 0.1)$. The spatial density that corresponds to one of these minima, showing its characteristic pear shape, is also plotted in Fig.~\ref{Fig1}. Around these minima the PES is rather soft in the interval $\beta_{3}\in[-0.2,+0.2]$. 
Moreover, a secondary minimum ($\sim4$ MeV above) is found at $(\beta_{2},\beta_{3})=(-0.1,0)$. The existence of the octupole deformed minima is a consequence of the location of certain single particle levels around the Fermi level, in particular, the proton $1h_{11/2}-2d_{5/2}$ and the neutron $1i_{13/2}-2f_{7/2}$ (see, e.g., Fig. 3 of Ref.~\cite{Egi.92}).

We have also drawn in Fig.~\ref{Fig1} the paths followed by two one-dimensional constrained calculations, i.e., the energy obtained when $\beta_{2}$ (dashed line) or $\beta_{3}$ (dotted line) is the only collective constrained variable in the HFB equation. In the case where only $\beta_{2}$ is considered we observe a collective path along the parity-conserving ($\beta_{3}=0$) direction in the oblate part ($\beta_{2}<0$) and a spontaneous parity symmetry-breaking in the prolate part ($\beta_{2}\in[0.0,+0.4]$), where the absolute minimum is found. On the other hand, only prolate deformed intrinsic states in the range ($\beta_{2}\in[+0.20,+0.35]$) are obtained by constraining only in $\beta_{3}$ -and leaving free the value of the quadrupole. In fact, these two deformations are correlated in this case, having larger values of $\beta_{2}$ for larger values of $\beta_{3}$.

The next step in the calculation of the spectrum of the nucleus $^{144}$Ba is the simultaneous restoration of the particle number, parity and angular momentum quantum numbers. In Fig.~\ref{Fig2} we plot the PESs that correspond to $J^{\pi}=0^{+},2^{+},1^{-}$ and $3^{-}$, where the reflection symmetry about the $\beta_{3}=0$ line is obtained again. We observe first that the absolute minima of the surfaces are located almost at the same $(\beta_{2},\pm\beta_{3})$ values as in the mean-field case. However, the potential wells around those points are a bit wider both in $\beta_{2}$ and $\beta_{3}$. For $J^{\pi}=0^{+}$, around $\sim4.5$ MeV of correlation energy is gained by the restoration of the symmetries at the absolute minima. Moreover, prolate and oblate minima are now connected through the $\beta_{3}$ degree of freedom since the barrier through the spherical point is much higher now. For $J^{\pi}=2^{+}$, prolate minima are much lower in energy than the oblate parity-symmetric ($\beta_{3}=0$) minimum. The same happens for $J^{\pi}=1^{-}$ and $3^{-}$ where the prolate minima are well-separated from the rest of the surface. We have to point out that for the intrinsic states with $\beta_{3}=0$, projection to odd-angular momenta and negative parity is not possible. The same happens at the spherical point $(\beta_{2},\beta_{3})=(0,0)$ for even-$J\neq0$. In those cases, a white band (point) in the PES shown in Fig.~\ref{Fig2} represents states with projected norm equal to zero.
\begin{figure}[t]
\begin{center}
\includegraphics[width=\columnwidth]{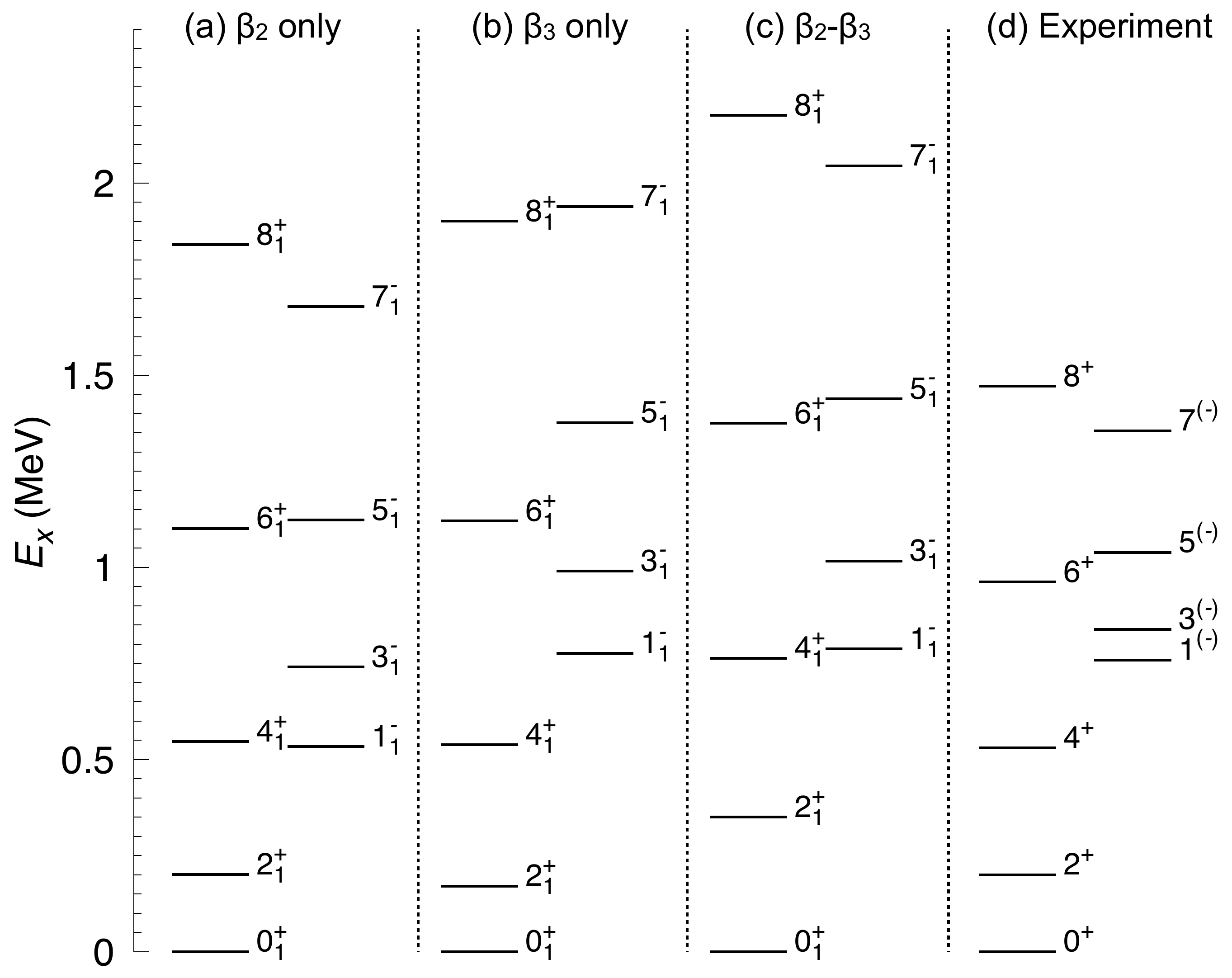}
\end{center}
\caption{(color online) Spectra (first two bands) obtained from SCCM calculations with (a) $\beta_{2}$, (b) $\beta_{3}$ and (c) $\beta_{2}-\beta_{3}$ as generating coordinates for $^{144}$Ba. Experimental data (d) are taken from Ref.~\cite{DataBase}}. 
\label{Fig3}
\end{figure}

Once the symmetries are restored, the final stage is shape mixing within the generator coordinate method. We have performed three different GCM calculations, namely, two one-dimensional GCM with $\beta_{2}$ and $\beta_{3}$ only as the collective coordinates, and one two-dimensional GCM with $(\beta_{2},\beta_{3})$ simultaneously taken into account. In the first two cases the states chosen to mix are those given by the collective paths marked in Fig.~\ref{Fig1}. Energies of the lowest positive and negative parity bands obtained by solving the corresponding HWG equations in these three different SCCM calculations are compared to the experimental spectrum in Fig.~\ref{Fig3}~\cite{FN}. In both 1D-GCM calculations we observe similar positive parity bands with a rotational character. However, the negative parity band obtained with $\beta_{3}$ (Fig.~\ref{Fig3}(b)) as the collective coordinate is globally shifted to higher excitation energies with respect to the one found with $\beta_{2}$ (Fig.~\ref{Fig3}(a)), keeping almost the same spacing between the levels. Including on the same footing axial quadrupole and octupole degrees of freedom (Fig.~\ref{Fig3}(c)) we observe a stretching of the positive parity band and a negative parity band similar to the one obtained with $\beta_{3}$ as the collective coordinate. Nevertheless, the absolute energies of the yrast states obtained in the 2D-GCM calculations are significantly the lowest among the three SCCM calculations, showing that the 2D-GCM calculation is better from the variational point of view. In particular, ground state energies calculated with 1D-GCM-$\beta_{2}$, 1D-GCM-$\beta_{3}$ and 2D-GCM-$(\beta_{2},\beta_{3})$ are $-1185.931$, $-1186.709$ and $-1187.547$ MeV, respectively.    

Comparing the theoretical results with the experimental data we observe first that only the 2D-GCM calculations can reproduce the relative position of the experimental levels (Fig.~\ref{Fig3}(d)). For instance, in the 1D-GCM results, the $1^{-}$ state is below the $4^{+}$ state in Fig.\ref{Fig3}(a) and the $7^{-}$ is above the $8^{+}$ in Fig.~\ref{Fig3}(b). Additionally, although the $2^{+}$ and $4^{+}$ energies are reasonably well reproduced by these two 1D-calculations, these bands have a stronger rotational character than the experimental one and the agreement with the experiment is lost for larger values of $J^{+}$. Finally, the experimental negative parity band is more compressed than those obtained with the present calculations and the energy of the band head state is well reproduced both by the 1D-GCM-$\beta_{3}$ and 2D-GCM-$(\beta_{2},\beta_{3})$ calculations. It is important to point out that a fully quantitative agreement with the experimental data cannot be expected within the present framework because neither triaxial ($K$-mixing) nor time-reversal symmetry breaking (cranking) intrinsic wave functions are considered. As a consequence, the ground state is better explored variationally than the excited states and gains more correlation energy producing the stretching of the spectrum. Including triaxial cranking intrinsic states would thus produce a compression of the calculated spectrum, and a better quantitative agreement with the experiments~\cite{Rod.00,PLB_746_341_2015,PRL_116_052502_2016}. However, these major improvements of the SCCM method are out of the scope of the present work.


\begin{table}
\begin{tabular}{cccccc} \hline\hline
$J^\pi_i \rightarrow J^\pi_f$ & $E\lambda$ & GCM $\beta_{2}$ & GCM $\beta_{3}$ & GCM $\beta_{2}-\beta_{3}$ & Exp \\ \hline\hline

$0^+\rightarrow2^+$  & E2 & 1.148 & 1.121 & 1.023 & 1.042$^{+17}_{-22}$ \\
$2^+\rightarrow4^+$  & E2 & 1.865 & 1.803 & 1.845 & 1.860$^{+86}_{-81}$ \\
$4^+\rightarrow6^+$  & E2 & 2.371 & 2.287 & 2.360 & 1.78$^{+12}_{-10}$  \\
$6^+\rightarrow8^+$  & E2 & 2.800 & 2.696 & 2.793 & 2.04$^{+35}_{-23}$  \\ \hline
$0^+\rightarrow1^-$  & E1 & 0.007 & 0.006 & 0.008 &       \\
$1^-\rightarrow2^+$  & E1 & 0.005 & 0.009 & 0.006 &       \\ \hline
$0^+\rightarrow3^-$  & E3 & 0.450 & 0.477 & 0.460 &  0.65$^{+17}_{-23}$ \\
$1^-\rightarrow4^+$  & E3 & 0.599 & 0.635 & 0.695 &       \\
$2^+\rightarrow5^-$  & E3 & 0.708 & 0.745 & 0.810 & $<1.2$ \\
$3^-\rightarrow6^+$  & E3 & 0.804 & 0.865 & 0.810 &      \\
$4^+\rightarrow7^-$  & E3 & 0.887 & 0.945 & 1.031 & $<1.6$ \\\hline
\end{tabular}
\caption{Absolute values of the transition matrix elements $|\langle J^\pi_i||E\lambda||J^\pi_f\rangle |$ (in $eb^{\lambda/2}$) for several transitions of interest. The experimental values are taken from \cite{PRL_116_052502_2016}.\label{tab:Table}}
\end{table}

The main advantage of the wave functions projected to good angular momentum is that they allow  a precise calculation of electromagnetic transition strengths without assuming the validity of the rotational approximation often used to relate intrinsic multipole moments with those transition strengths. The rotational approximation is valid in the strong deformation limit, which is not reached for many of the relevant configurations in the present calculation. This might lead to substantial qualitative deviations in the evaluation of transition strengths~\cite{PRC_86_054306_2012}. In table~\ref{tab:Table} we compare our SCCM results (1D and 2D) for the absolute value of the transition strengths $|\langle J^\pi_i||E\lambda||J^\pi_f\rangle|$ for selected transitions with the available experimental data. We observe first that the results with 1D and 2D calculations are rather similar for this nucleus. In addition, for the positive parity rotational band, the in-band $E2$ transitions follow rather well the rotational behavior and agree very well with experimental data for the two lowest transitions. At higher spins, the experimental data deviate from the rotational behavior probably due to the quenching of pairing correlations that our calculations cannot reproduce. For a proper theoretical description we would need to carry out proper variation after projection (VAP) calculations that would lead to cranking type intrinsic states, a feat that is out of reach with the present computational capabilities. Nevertheless, for transitions to the ground state, the effect of including cranking terms is expected to be small as they are related to changes in the intrinsic deformations. For the $0^+\rightarrow 3^-$ transition, the theoretical prediction is smaller than the experimental value but within the error bars. Our result, expressed in Weisskopf units is $B(E3,3^-\rightarrow 0^+)=25 $ W.u. which is a bit too low as compared to the experimental value of $48^{+25}_{-34}$ W.u.~\cite{PRL_116_112503}. However, our result agrees well with systematic in the region \cite{ki02}. On the other hand, the $E1$ transitions are rather small, as a consequence of the small dipole moment of the intrinsic states in this nucleus~\cite{Egi.92}.

\begin{figure*}[t]
\begin{center}
\includegraphics[width=\textwidth]{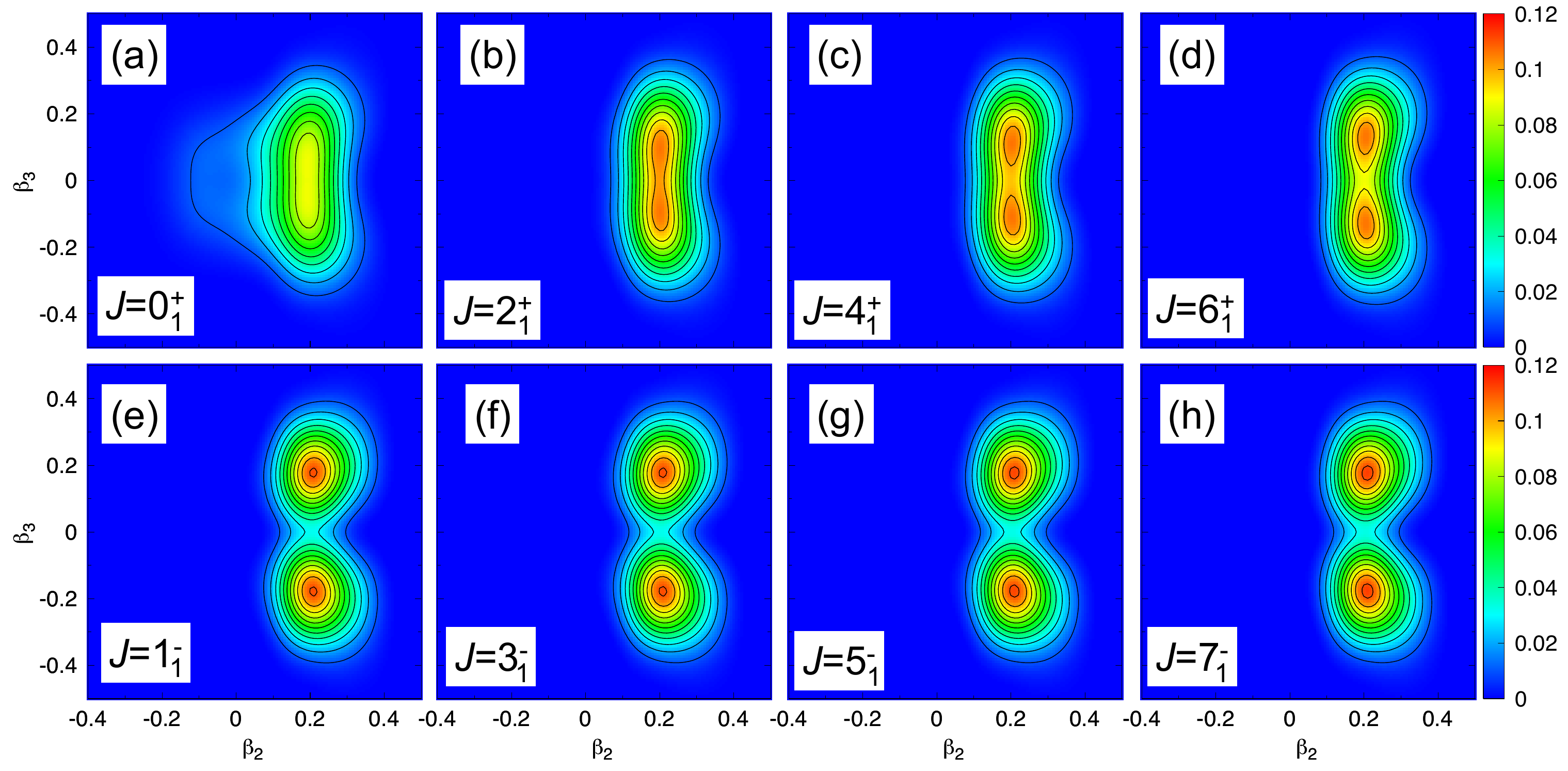}
\end{center}
\caption{(color online) Collective wave functions for the ground state (upper part) and first excited negative parity (lower part) bands.}
\label{Fig4}
\end{figure*}

In Fig.~\ref{Fig4} the square of the collective amplitudes in Eq.~\ref{coll_wf} are plotted for the lowest lying states of each angular momentum and parity. The amplitudes must be even (odd) functions under the $\beta_3\rightarrow -\beta_3$ reflection for even (odd) $J$ values. In the latter case, this implies that the amplitudes must vanish along the $\beta_3=0$ line. As a consequence of this restriction the negative parity amplitudes are shifted towards larger octupole moments. In Fig.~\ref{Fig4} we also observe that the members of the same band, which are strongly connected by electromagnetic transitions, share a similar structure of their collective wave functions, which is evident in the negative parity band. In the positive parity band we see an evolution of the intrinsic state with increasing spin that is associated to the stabilization of the octupole deformation in this case \cite{PRL_80_4398_1998}. As a consequence, the collective amplitudes for $J\geq 4^{+}$ peak at values of $\beta_3\neq0$.

To summarize, we have carried out state-of-the-art calculations with the Gogny D1S interaction to describe the lowest lying positive and negative parity states of $^{144}$Ba. Angular momentum, parity and particle number symmetries, broken by quadrupole and octupole constrained HFB states, have been restored and these projected states are allowed to mix within the GCM framework. The results for the excitation energies agree qualitatively with the experimental results when both quadrupole and octupole degrees of freedom are treated on an equal footing. A stretched theoretical spectrum is obtained due to the lack of triaxial and time-reversal symmetry breaking components. Their inclusion would bring the predictions closer to the experimental values. However, including these terms requires major developments of the method that are out of the scope of the present work. 
Finally, transition strengths are in a rather good agreement with the experimental data. The calculated $B(E3,3^{-}\rightarrow 0^{+})$ strength is predicted to be lower than the most recent measurements although it is within the large experimental error bar and agrees well the systematics. Future experiments would be very helpful to disentangle the actual amount of octupole correlations in $^{144}$Ba.    

\section*{Acknowledgements}
We acknowledge the support from GSI-Darmstadt and CSC-Loewe-Frankfurt computing facilities. This work was supported by the Ministerio de Econom\'ia y Competitividad under Contracts Nos. FIS2012-34479, FIS-2014-53434, FPA2015-65929 and FIS2015-63770 and Programa Ram\'on y Cajal 2012 No. 11420.



\begin{thebibliography}{10}
\bibitem{FN}
Please note that the GCM calculation with $\beta_2$ as collective coordinate produces meaningful results for negative parity states because the set of HFB generating states includes octupole deformed states. If it not were the case, zero norm negative parity states would have been obtained.
\end{thebibliography}
\end{document}